\documentstyle[pre,preprint,aps,epsfig]{revtex}

\begin{document}

\draft

\tighten

\title{Resonances in the dynamics of $\phi^4$ kinks perturbed by ac 
forces} 

\author{Niurka R.\ Quintero$^{*}$ and Angel S\'anchez$^{\dag}$} 

\address{Grupo Interdisciplinar de Sistemas
Complicados (GISC), Departamento de Matem\'aticas, 
Universidad Carlos III de Madrid,\\
Edificio Sabatini,
Avenida de la Universidad 30, E-28911 Legan\'{e}s, Madrid, Spain}

\author{Franz G.\ Mertens$^{\ddag}$}

\address{Physikalisches Institut, Universit\"at Bayreuth,  
D-95440 Bayreuth, Germany}

\date{\today}

\maketitle

\begin{abstract} 
We study the dynamics of 
$\phi^4$ kinks perturbed by an ac force, both with and without damping.
We address this issue by using a collective coordinate theory, which 
allows us to reduce the problem to the dynamics of the kink center and
width.
We carry out a careful analysis of the corresponding ordinary
differential equations, of Mathieu type in the undamped case, finding
and characterizing the resonant frequencies and the regions of existence
of resonant solutions.
We verify the accuracy of our predictions
by numerical simulation of the full partial differential equation, 
showing that the collective coordinate prediction is very accurate. 
Numerical simulations for the damped case establish that the strongest
resonance is the one at half the frequency of the internal mode of the
kink. 
In the conclusion we discuss on the possible relevance of our results
for other systems, especially the sine-Gordon equation.
We also obtain additional results regarding the equivalence
between different collective coordinate methods applied to this problem.
\end{abstract}


\narrowtext 

\section{Introduction}

\label{intro}

One century and a half after their discovery, solitons and solitary waves
have proven themselves ubiquitous in nature, arising in very many physical
applications and leading to very important advances in applied mathematics
\cite{Remo,Scott}. Generally speaking, it is often the case that the 
properties of solitary waves are known for certain equations, perhaps 
integrable, that relate to an oversimplified description of different 
physical systems; subsequently, one is interested to learn how these 
properties are modified if terms initially neglected are included as 
perturbations of the original equation. In most cases, this is a very 
complicated problem, and shedding light on it usually requires the use
of approximate analytical approaches. One of the most succesful and widely
applicable of these approaches is the collective coordinate technique or,
rather, the family of collective coordinate techniques \cite{yurirev,siam}.
The main merit of these procedures is the drastic reduction of the number
of degrees of freedom involved in the problem, from the infinity of them
in the original partial differential equation to the dynamics of a few
degrees of freedom, governed by ordinary differential equations. 
Quite commonly, the reduction is done to a single
degree of freedom, which, in general, can be identified with 
the center of the wave $X(t)$ or its velocity $V(t)$, as was  
first proposed in the mid-seventies \cite{Fogel,McL}.
This amounts to mapping the motion of solitons or 
solitary waves 
to the motion of a pointlike (perhaps relativistic) particle with 
an effective mass \cite{fdez}. Surprisingly,
this dramatic simplification leads
to excellent results for very many systems \cite{yurirev,siam}. 
However, there are perturbations which, when acting on solitary waves,
change not only the position of the center of mass of the kink, 
but also its width, as suggested in \cite{Rice},
or even some radiation can appear in the waves, a phenomenon that
can yield the whole collective coordinate idea useless as more and 
more degrees of freedom are excited \cite{fer1,maje}. 

As we have mentioned above, 
in 1983 Rice \cite{Rice} 
developed a new perturbative 
method, which he applied to two well-known nonlinear Klein-Gordon 
problems, the $\phi^4$ and the sine-Gordon equations \cite{Remo,Scott},
in order to account for variations of the width of their kink solutions
under perturbations. In Rice's approach, 
the collective coordinates are the kink center $X(t)$ and its 
width $l(t)$. His results point out that when, in those systems, the kink 
is subject to ``some perturbations'' ({\em sic}\/) the simple 
translational motion of the kink center can be coupled to an oscillatory 
motion of the width of the wave, whose frequency he obtained by means
of a variational approach. Interestingly, 
for the $\phi^4$ equation this so-called Rice frequency, $\Omega_R$, 
practically coincides with the frequency $\Omega_i$ 
of the kink internal mode (one of the modes
of linear excitations around the kink, corresponding to a nonzero 
eigenvalue in the discrete spectrum, see, e.g., \cite{Jey}). 
On the contrary, for the sine-Gordon (sG)
case, $\Omega_R$ turns out to be within the
phonon spectrum and, furthermore, the sG kink does not have an internal
mode (its only eigenvalue in the discrete spectrum is zero, corresponding
to a Goldstone mode \cite{Fogel}). 
For this reason, the
deformation of the kink width due to the internal mode has been studied 
extensively in the $\phi^4$ equation and, among other interesting 
results, we now know 
that the internal 
mode is able to store and transfer the energy 
in $\phi^4$ kink-antikink collisions \cite{campbell}, 
in the interaction of $\phi^4$ kinks with impurities \cite{Yura},
or in the case 
when the $\phi^4$ kink is subject to a periodic spatially modulated potential 
\cite{fei}. This exchange of energy between the internal and the 
translational modes  
or among the internal mode and the modes of the impurities 
explains the resonances that take place in the above perturbed 
$\phi^4$ systems. 

In view of these results, a question that naturally arises is the possible
existence of resonances of the $\phi^4$ kink when perturbed by
external ac forces, a problem that has not been considered so far and 
that relates to a number of physical contexts. Very recently, we have 
shown \cite{prl} that a strong resonance arises when the $\phi^4$ system is 
subject to an external ac force of the form $f(t) = 
\epsilon \sin(\delta t + \delta_{0})$ 
with $\delta$ 
close to $\Omega_{i}/2$, while 
in the case that $\delta = \Omega_{i}$ this resonance is weak and
even disappears  
for an appropriate choice of the initial kink velocity and the parameters 
of the driving force. However, the work reported in \cite{prl} contained 
mostly numerical results and only an intuitive explanation of this striking
phenomenon in terms of the collective coordinate equations, and therefore 
our aim in this paper is to provide a full analytical treatment of those
equations (for the undamped case) in order to better understand the anomalous
resonance phenomenon. In addition, in doing so we will be already carrying 
out the same analysis for the sG equation, which may be of interest in order
to clarify the question of the existence of an internal quasi-mode for this
equation \cite{willis}. We deal with these issues along the paper 
according to the following scheme: 
In the next Section we use the so called 
generalized travelling wave {\em Ansatz} (GTWA) \cite{franz}
to obtain the equations governing the dynamics of the kink center and
width (which is associated to the excitation of the internal mode). 
In Sec.\ III we thoroughly analyze those equations in the absence of 
damping, identifying all the possible resonances and their locations 
as a function of the equation parameters. We then focus on the most 
interesting resonances, namely those at $\Omega_i/2$ and $\Omega_i$. 
For the damped case we have not been able to solve analytically the equation for 
$l(t)$, but in Sec.\ IV we present numerical simulations of the full 
partial differential equation confirming 
that the resonance at $\Omega_{i}/2$ remains for weak damping, while 
the resonance at  $\delta \approx \Omega_{i}$ disappears for any 
damping value. Additionally, in that section
we have compared all our analytical results 
to the numerical simulations, finding an excellent agreement.  
Finally, in the last section we summarize our results and discuss their
implications for other systems, focusing especially on the sG kink 
dynamics. In an appendix, we prove the equivalence of the GTWA to 
a different procedure to obtain collective coordinate equations, which uses
the momentum and the energy of the system similarly to the classical
approach in \cite{McL}. 

\section{Collective coordinate approach}

The dynamics of $\phi^4$ kinks subject to 
a periodic force  $f(t)=\epsilon \sin(\delta t + \delta_{0})$ 
is governed by the equation
\begin{equation}
\begin{array}{c}
{\displaystyle{
\phi_{tt} - \phi_{xx} - \phi + \phi^3 = -\beta \phi_{t} + f(t)}},
\label{ecua1}
\end{array}
\end{equation}
where $\beta$ is the damping coefficient, and $\epsilon$, $\delta$ and 
$\delta_{0}$ represent the amplitude, the frequency and the phase of the 
periodic force, respectively.

In order to apply the GTWA, first proposed in \cite{franz}, we rewrite
Eq.\ (\ref{ecua1}) in the following way: 
\begin{eqnarray}
\label{ecua6}
\dot{\phi}& =& \frac{\delta {\it {H}}} {\delta \psi},\\
\dot{\psi}&  = & -\frac{\delta {\it {H}}} {\delta \phi} + 
F(x, t, \phi, \phi_{t}, ...), 
\label{ecua7}
\end{eqnarray}  
where $\psi = \dot{\phi}$ and the dot represents derivative 
with respect to time, $F(x, t, \phi, \phi_{t}, ...)=-\beta \dot{\phi} + f(t)$, 
 and $H$ is the Hamiltonian of the system with 
the corresponding 
potential, ${\displaystyle{ U(\phi)=  (\phi^2-1)^2/4}}$,  
when 
$\epsilon$ and $\beta$ are zero,  
\begin{equation}
\begin{array}{l}
{\displaystyle{ 
{\it{H}} = \int_{-\infty}^{+\infty} dx \Big \{ 
\frac{1}{2} \psi^2 + \frac{1}{2} \phi_{x}^{2} + U(\phi) \Big\}}}.
\end{array}
\label{ecua8}
\end{equation}
We now assume that the
solution of (\ref{ecua6})-(\ref{ecua7}) has the form  
\begin{equation}
\begin{array}{l}
{\displaystyle{
\phi(x,t) = \phi[x - X(t), l(t)]}},
\end{array}
\label{ecua9}
\end{equation}
and, hence, from the definition of $\psi$ we have that 
\begin{equation}
\begin{array}{l}
{\displaystyle{ 
\psi(x,t) = \psi \left[x - X(t), l(t), \dot{X}, \dot{l}\right]}}.
\end{array}
\label{ecua9a}
\end{equation}
Here $\phi$ describes the soliton shape, whose center will be given
by $X(t)$, and where 
we introduced a second collective variable, $l(t)$, which 
will represent the kink width as we will see below. This {\em Ansatz} takes 
into account that
$l$ can be different from the Lorentz-contracted width due to the action of
the external force. 
Since the internal mode is related with the kink width,
we can expect that by using this approach we will be able to explain 
the kink motion when the internal mode is excited. However, the
linearized  problem around the initial kink solution 
tells us that perturbations 
of the kink can not only shift its position or change its width, but 
also make it emit radiation \cite{Jey}. 
We have not considered this effect 
and therefore the {\em Ansatz} is only valid when the external force 
is so small that practically no radiation is emitted. 
Since the frequencies of the discrete internal 
mode $\Omega_{i} = \sqrt{3/2}$ and the continuum phonon band 
$\omega_{p} = \sqrt{2 + k^2/2}$ are apart from each other one can expect that  
the two kinds of 
modes are excited at different values of the parameters of the external 
force, and then when resonances appear as a consequence of the excitation
of the internal 
mode one can in principle neglect the effect of radiation.
In any event all our analytical results will have to be confirmed later
by numerical simulations, which will show whether or not this assumption
is correct.

To obtain the equations for our 
collective coordinates $X(t)$ and $l(t)$ we will follow \cite{franz} 
(see the appendix for another possible derivation). 
First, we insert (\ref{ecua9})-(\ref{ecua9a}) 
into (\ref{ecua6}) and 
(\ref{ecua7}), and then we multiply the first obtained equation 
by $\partial \psi/\partial X$ and the second one 
by  $\partial \phi/\partial X$;
subtracting both expressions
and integrating we arrive at the following system of equations:
\begin{equation}
\begin{array}{l}
{\displaystyle{
\int_{-\infty}^{+\infty} dx \,\ \frac{\partial \phi}{\partial X} 
\frac{\partial \psi}{\partial \dot{X}} \,\ \ddot{X} +   
\int_{-\infty}^{+\infty} dx \,\  [\phi, \psi] \,\ \dot{l} + 
\int_{-\infty}^{+\infty} dx \,\  \frac{\partial \phi}{\partial X} 
\frac{\partial \psi}{\partial \dot{l}} \,\ \ddot{l} - F^{stat} (X) =}}
  \nonumber \\
{\displaystyle{ 
 \,\ = \,\ \int_{-\infty}^{+\infty} dx F(x,t,\phi,\phi_{t},...) 
\frac{\partial \phi}{\partial X}}}, 
\end{array}
\label{ecua11}
\end{equation}
with
\begin{equation}
\begin{array}{l}
{\displaystyle{  
[\phi, \psi] =   
 \frac{\partial \phi}{\partial X} \frac{\partial \psi}{\partial l} - 
 \frac{\partial \phi}{\partial l} \frac{\partial \psi}{\partial X},}}
\end{array}
\label{ecua12}
\end{equation}
\begin{equation}
\begin{array}{l}
{\displaystyle{ 
F^{stat} = -\int_{-\infty}^{+\infty} dx \,\ \left \{  
\frac{\delta {\it{H}}}{\delta \phi} \frac{\partial \phi}{\partial X}  +  
\frac{\delta {\it{H}}}{\delta \psi} \frac{\partial \psi}{\partial X} 
\right \} = 
 -\int_{-\infty}^{+\infty} dx \,\ \frac{\partial {\cal{H}}}{\partial X} = 
- \frac{\partial E}{\partial X}}}, 
\end{array}
\label{ecua13}
\end{equation}
where $E$ represents the energy of the system, $\cal{H}$ is the 
Hamiltonian density and $F^{stat}$ is the static force due an 
the external field or other solitons, 
equal to zero for the above Hamiltonian. 
In order to obtain the second equation of motion we proceed as in 
the previous equation and begin by inserting (\ref{ecua9}) and (\ref{ecua9a}) 
into (\ref{ecua6}) 
and (\ref{ecua7}), then multiplying the first and second equation by 
 $\partial \psi/\partial l$ and 
 $\partial \phi/\partial l$ respectively,  and finally
taking the difference and integrating, thus finding 
\begin{equation}
\begin{array}{l}
{\displaystyle{
\int_{-\infty}^{+\infty} dx \,\  [\psi, \phi] \,\ \dot{X}  + 
\int_{-\infty}^{+\infty} dx \,\ \frac{\partial \phi}{\partial l} 
\frac{\partial \psi}{\partial \dot{X}} \,\ \ddot{X} + 
\int_{-\infty}^{+\infty} dx \,\  \frac{\partial \phi}{\partial l} 
\frac{\partial \psi}{\partial \dot{l}} \,\ \ddot{l} - K^{int} (X)  =}}
\\ 
{\displaystyle{
 \,\ = \,\ \int_{-\infty}^{+\infty} dx \,\ F(x,t,\phi,\phi_{t},...) 
\frac{\partial \phi}{\partial l}}}, 
\end{array}
\label{ecua15}
\end{equation}
where 
\begin{equation}
\begin{array}{l}
{\displaystyle{
K^{int}(l,\dot{l},\dot{X}) \,\ = \,\ - \int_{-\infty}^{+\infty} dx \,\ 
\frac{\partial {\cal{H}}}{\partial l} = 
- \frac{\partial E}{\partial l}}}. 
\end{array}
\label{ecua16}
\end{equation}
By this procedure,
we have obtained two coupled second-order ordinary differential
equations for $X(t)$ and $l(t)$, where 
up to now we have not imposed any condition on the soliton shape; 
however, to  solve Eqs.\
(\ref{ecua11})--(\ref{ecua16}) we need the explicit functional 
dependence of  
$\phi(x,t)$. Following Rice \cite{Rice} we assume
that 
\begin{equation}
\begin{array}{l}
{\displaystyle{ 
\phi(x,t) = \phi_{0} [x-X(t), l(t)] = 
\tanh \left [\frac{x - X(t)}{ l(t)} \right],  
}}
\end{array}
\label{ecua17}
\end{equation}
where ${\displaystyle{\phi_{0}=}=\tanh[x-X_{0}/l_{0}]}$ 
is the static kink solution of the $\phi^{4}$ 
system, centered at $X_{0}$ and of width $l_{0}=\sqrt{2}$.
Substituting (\ref{ecua17}) into  (\ref{ecua11})--(\ref{ecua16}) and 
integrating over $x$ we obtain 
\begin{eqnarray}
\label{ecua18}
M_{0} l_{0} \frac{\ddot X}{l} - M_{0} l_{0} \frac{\dot X \dot l}{l^{2}} 
 &=&  F^{stat}(X) -  \beta M_{0} l_{0} \frac{\dot X }{l} + F_{ex}, \\
\alpha M_{0} l_{0} \frac{\ddot l}{l} + M_{0} l_{0} \frac{\dot X^{2}}{l^{2}} 
 &=& K^{int}(l,\dot{l},\dot{X}) -  
\beta \alpha M_{0} l_{0} \frac{\dot l}{l} + K, 
\label{ecua19}
\end{eqnarray}
where  
\begin{equation}
\begin{array}{l}
{\displaystyle{
F_{ex}  =  \int_{-\infty}^{+\infty} dx \,\ f(t) 
\frac{\partial \phi}{\partial X} = - q f(t), \,\ \,\ F^{stat}=0}}, 
\end{array}
\label{ecua20}
\end{equation}
\begin{equation}
\begin{array}{l}
{\displaystyle{
K = \int_{-\infty}^{+\infty} dx \,\ f(t) \,\ 
\frac{\partial \phi}{\partial l}=0, \,\ \,\ 
K^{int} = -\frac{\partial E}{\partial l}}},
\end{array}
\label{ecua21}
\end{equation}
\begin{equation}
\begin{array}{l}
{\displaystyle{
E = \frac{1}{2} \frac{l_{0}}{l} M_{0}  \dot{X}^{2} + 
\frac{1}{2} \frac{l_{0}}{l} \alpha M_{0}  \dot{l}^{2} + 
\frac{1}{2} M_{0} \left(\frac{l_{0}}{l} + \frac{l}{l_{0}}\right)}}, 
\end{array}
\label{ecua22}
\end{equation}
with $\alpha=(\pi^2-6)/12$, $q=2$ and $M_0=4/(3l_0)$.   
Denoting $P(t) \equiv M_{0} l_{0} \dot{X}/l$, the Eq.\ (\ref{ecua18}) 
can be written as 
\begin{equation}
\begin{array}{l}
{\displaystyle{
\frac{dP }{dt} = -  \beta P - q f(t)}}.  
\end{array}
\label{ecua23}
\end{equation}
It is interesting to note that this 
equation may be obtained as well 
by applying the McLaughlin and Scott procedure \cite{McL}
with one collective variable only corresponding to the center of the 
kink.  As was shown in \cite{EPJBus}, for the sG kink dynamics this 
is an excellent description of the kink motion, and we have verified 
that it also describes the
$\phi^4$ kink motion under ac forces away from the resonances we will 
find and discuss below. Its solution is given by 
\begin{equation}
\begin{array}{c}
{\displaystyle{
P(t)  =
 \frac{q \,\ \epsilon}{(\beta^{2} +\delta^{2})}
\left[\delta \cos(\delta t + \delta_{0})
-\beta \sin(\delta t + \delta_{0})
 \right] +}} \\ \nonumber
{\displaystyle{+ \exp(-\beta t) \left \{ P(0) +
\frac{q \,\ \epsilon} {(\beta^{2} +\delta^{2})}
\left[\beta \sin(\delta_{0}) -
\delta \cos(\delta_{0}) \right] \right \}}}.
\end{array}
 \label{ecua4}
\end{equation}   
{}From Eq.\ (\ref{ecua19}) the equation that holds for the width $l(t)$ is 
\begin{equation}
\begin{array}{l}
{\displaystyle{
\alpha \left[\dot{l}^{2} - 2 l \ddot{l} - 2 \beta l \dot{l}\right]  =  
\frac{l^{2}}{l_{0}^{2}} \left [1 +  \frac{P^{2}}{M_{0}^{2}} \right ] - 1}}. 
\end{array}
\label{ecua24}
\end{equation}
Note that the term $P(t)^2$ in Eq.\ (\ref{ecua24}) involves two 
frequencies $\delta$ and $ 2 \delta$ when $\beta = 0$; whereas for 
$\beta \ne 0$ and after 
some transient time the only frequency that remains is $2 \delta$. 
Furthermore, the term with frequency $\delta$ vanishes for $\beta=0$ and 
an appropriate choice of initial parameters.  
This equation represents a nonlinear, damped and parametrically excited 
oscillator. 
To solve Eq.\ (\ref{ecua24}) we provide the following change of 
variables, proposed 
in \cite{yuri}, $l(t)=g^{2}(t)$, which transforms the above equation into 
an Ermakov-type equation (or Pinney-type) 
[see \cite{reid} and references therein]. 
Then,  
the equation for $g(t)$ reads  
\begin{equation}
\begin{array}{l}
{\displaystyle{
\ddot{g} + \beta \dot{g}  
+ \left[ \left( \frac{\Omega}{2} \right)^2 + 
\left( \frac{\Omega}{2 M_{0}} \right)^2 P^2 \right ] g =  
\frac{1}{4 \alpha g^{3}}}}, \nonumber \\
{\displaystyle{
 g(0) = \sqrt{l_{s}} \ne 0,   \,\ \,\ \,\   
\dot{g}(0) = \frac{\dot{l}(0)}{2 \sqrt{l_{s}}}}},
\end{array}
\label{ecua25}
\end{equation}
where $\Omega = 1/\sqrt{\alpha}  l_{0}=1.2452$ is equal to   
the Rice frequency $\Omega_{R} = 1/\sqrt{\alpha}  l_{s}$ in the case when the 
kink initially is at rest; this agrees within $1.7\%$ with 
$\Omega_{i}=\sqrt{3/2}=1.2247$.   
We have not been able to solve analytically Eq.\ (\ref{ecua25}), 
except when $\beta = 0$. The next section is devoted to a detailed analysis
of that case; we will come back to the nonzero $\beta$ problem when discussing
our numerical results in Sec.\ IV. 

\section{Undamped kink: $\beta = 0$}
 
When $\beta=0$, Eq.\ (\ref{ecua25}) reads   
\begin{equation}
\begin{array}{c}
{\displaystyle{
\ddot{g} + 
\left [ \left(\frac{\Omega}{2}\right)^2 + 
\left(\frac{\Omega}{2 M_{0}}\right)^2 P^2 \right ] g =  
\frac{1}{4 \alpha g^{3}}}}, 
\end{array}
\label{ecua26}
\end{equation}
where $P(t)$ is given by 
${\displaystyle{ 
P(t)= \lambda +q \epsilon\cos(\delta t + \delta_{0})/\delta
}}$, with
${\displaystyle{  
\lambda \equiv M_{0}  \gamma_{0} u(0)/l_{s}  
- q \epsilon \cos(\delta_{0})/\delta }}$. We thus see that
the function $P(t)^2$ in Eq.\ (\ref{ecua26}) 
involves trigonometric functions with frequencies $\delta$ and $2 \delta$
if and only if 
${\displaystyle{ \lambda \ne 0 }}$. 
On the contrary, when $\lambda=0$ the only frequency that remains in
the function $P(t)^2$ is $2 \delta$. 
Interestingly, we note that the
relation $\lambda=0$ coincides with the condition for the 
oscillatory motion of the center 
of the kink, obtained by using the McLaughlin and Scott 
approach in the absence of dissipation \cite{EPJBus}.
The solution of Eq.\ (\ref{ecua26}) \cite{Pinney} is 
\begin{equation}
\begin{array}{c}
{\displaystyle{
g(t)  =  \sqrt{v_{1}^{2} + \frac{1}{4 \alpha W^{2}} v_{2}^{2}}}}, 
\end{array}
\label{ecua27}
\end{equation}
where $v_{1}(t)$ and $v_{2}(t)$ are two independent solutions of the linear
part of Eq.\ (\ref{ecua26}), $W=\dot{v}_{1} v_{2} - \dot{v}_{2} v_{1}$ 
is the Wronskian, 
which in this case  is a constant and can be calculated by 
the initial conditions for
$v_{i}$ ($i=1,2$) which are $v_{1}(0)=\sqrt{l_{s}}$,  $\dot{v}_{1}(0) 
= \dot{l}(0)/(2 \sqrt{l_{s}})$, 
$v_{2}(0) = 0$ and  $\dot{v}_{2}(0) = const \ne 0$. 

\subsection{$\lambda = 0$: Resonance at $\delta \approx \Omega_{R}/2$}

\begin{figure}
\begin{tabular}{c}
\epsfig{file=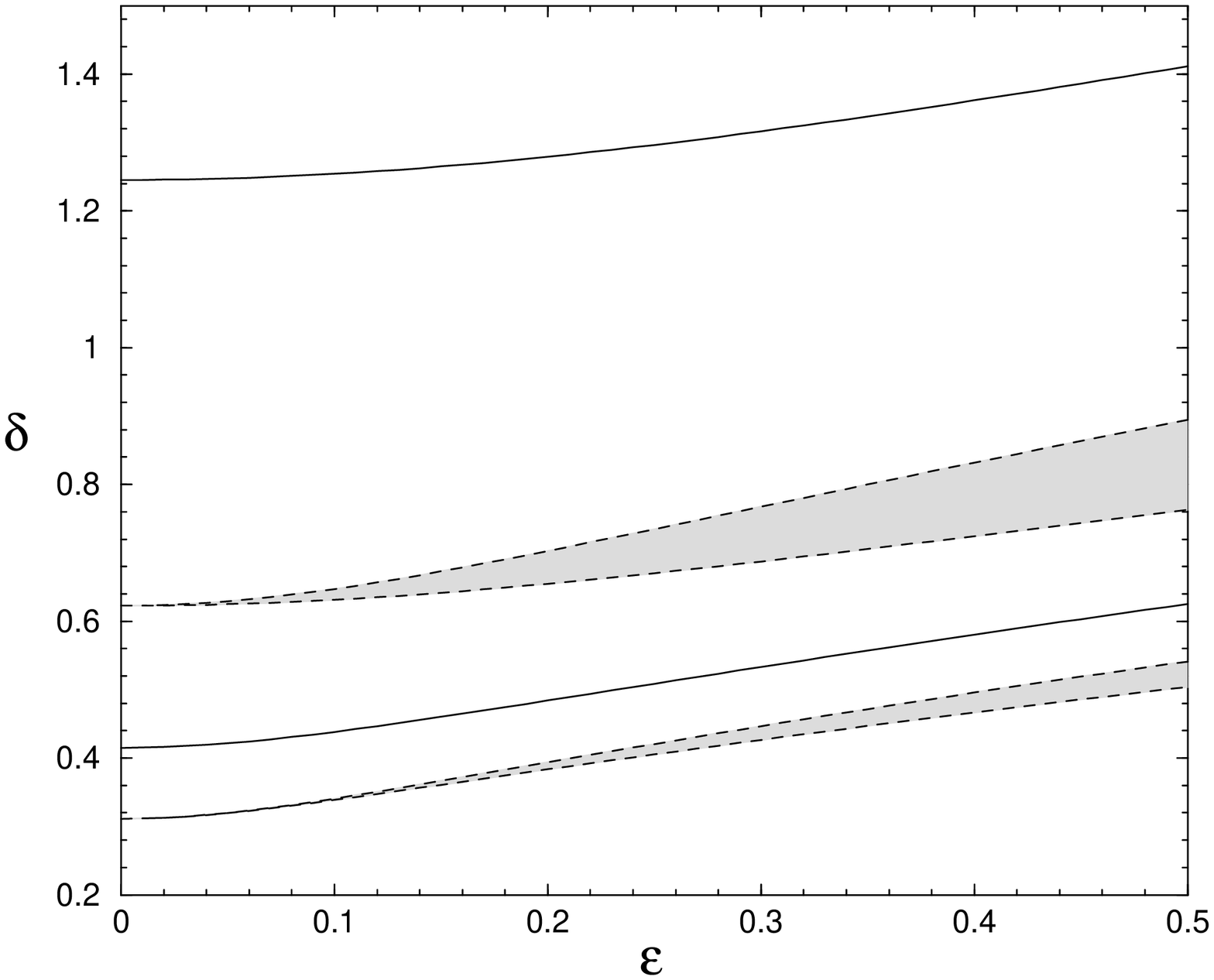, width=3.0in, angle=-90}
\end{tabular}
\caption{Characteristic curves $\delta=\delta(\epsilon)$ corresponding to 
Eqs.\ (\ref{ecua36}) [upper solid line], (\ref{ecua37}) and (\ref{ecua38}) [upper dashed lines] 
for zero initial kink velocity and $\delta_{0}=\pi/2$. 
The lower solid line is related with the solution (\ref{ecua32})-(\ref{ecua35}) 
when $\nu=3/2$. The lower dashed 
lines are the characteristic curves of the
integer Mathieu functions with $n=2$. 
Shadowed regions represent unstable solutions of Eq.\ (\ref{ecua28}).  
}
\label{f2}
\end{figure}

If one denotes $\tau = \delta t + \delta_{0}$ and sets $\lambda=0$ in the 
linear part of Eq.\ (\ref{ecua26}), after some manipulations    
we arrive at the following Mathieu  equation for the $v_{i}$ functions:
\begin{equation}
\begin{array}{c}
{\displaystyle{
v_{i}'' + [a + 2 \theta \cos(2 \tau) ] v_{i} = 0}},  \nonumber \\
{\displaystyle{
a=\left(\frac{\Omega}{2 \delta}\right)^2  \left[1 + \frac{q^2 \epsilon^2}
{2 \delta^2 M_{0}^2} \right], \,\ \,\ 
\theta = \left(\frac{\Omega}{2 \delta}\right)^2 
\frac{q^{2} \epsilon^2}{4 \delta^{2} M_{0}^2}}}, 
\end{array}
\label{ecua28}
\end{equation}
where prime denotes the derivative with respect to $\tau$. 
Notice that the initial  conditions for $v_{i}(\tau)$ 
become $v_{1}(\delta_0)=\sqrt{l_{s}}$,  
$v_{1}'(\delta_0)= \dot{l}(0)/(2 \delta \sqrt{l_{s}})$,  
$v_{2}(\delta_0) = 0$ and $v_{2}'(\delta_0)= \dot{v}(0)/\delta$. 
The solution of Eq.\ (\ref{ecua28}) (see \cite{Mathieu}) for $v_{1}(\tau)$ and
$v_{2}(\tau)$ can be expressed as  a linear  superposition of the two Mathieu
functions $ce_{\nu}$ and $se_{\nu}$ 
with a non-integer index $\nu$, i.e., 

\begin{equation}
\begin{array}{l}
{\displaystyle{
v_{i}(\tau) = A_{i} \, {\rm {ce}}_{\nu}\left(\tau,-\theta\right) + 
B_{i} \, {\rm{se}}_{\nu}\left(\tau,-\theta\right), \,\ \,\ i=1,2,}}  
\end{array}
\label{ecua29}
\end{equation}
where
\begin{equation}
\begin{array}{l}
{\displaystyle{
 A_{i}  \equiv  \frac{\Delta_{A_{i}}}{\Delta}, \,\ \,\ 
 B_{i} \equiv \frac{\Delta_{B_{i}}}{\Delta}}},  
\label{ecua30} 
\end{array}
\end{equation}
and
\[
\begin{array}{l}
{\displaystyle{
\Delta = {\rm {ce}}_{\nu}\left(\delta_{0},-\theta\right) \, 
 {\rm {se}}'_{\nu}\left(\delta_{0},-\theta\right) \, - \,
{\rm {ce}}'_{\nu}\left(\delta_{0},-\theta\right) \,
 {\rm {se}}_{\nu}\left(\delta_{0},-\theta\right)}}, \nonumber \\[3mm] 
{\displaystyle{
\Delta_{A_{i}} = v_{i}(\delta_{0}) \, 
{\rm {se}}'_{\nu}\left(\delta_{0},-\theta\right) \, - \,  
v'_{i}(\delta_{0}) \,
 {\rm {se}}_{\nu}\left(\delta_{0},-\theta\right)}}, \nonumber \\[3mm]
{\displaystyle{
\Delta_{B_{i}} = v'_{i}(\delta_{0}) \,  
{\rm {ce}}_{\nu}\left(\delta_{0},-\theta\right) \, - \,  
v_{i}(\delta_{0}) \, {\rm {ce}}'_{\nu}\left(\delta_{0},-\theta\right)}} 
, \nonumber
\end{array}
\]
with the constraint (characteristic curve for Mathieu functions)
\begin{equation}
\begin{array}{l}
{\displaystyle{
a = \nu^{2} + \frac{1}{2 (\nu^{2}-1)} \theta^{2} + O(\theta^{4})}}. 
\end{array}
\label{ecua31}
\end{equation}
{}From Eqs.\ (\ref{ecua27}), (\ref{ecua29}) and (\ref{ecua30}), and taking
into  account that $\tau = \delta t + \delta_{0}$ 
we obtain that the kink width $l(t)$
is given by 
\begin{equation}
\begin{array}{l}
{\displaystyle{
l(t) = g^{2} = v_{1}^{2}(t) + \frac{1}{4 \alpha W^{2}} v_{2}^{2}(t)}}, 
\end{array}
\label{ecua32}
\end{equation}
\begin{equation}
\begin{array}{l}
{\displaystyle{
v_{i}(t) = A_{i} \,  
{\rm {ce}}_{\nu}\left(\delta t + \delta_{0},-\theta\right) + 
B_{i} \, {\rm{se}}_{\nu}\left(\delta t + 
\delta_{0},-\theta\right), \,\ \,\ i=1,2}}; 
\end{array}
\label{ecua33}
\end{equation}
where 
\begin{equation}
\begin{array}{l}
{\displaystyle{
 A_{i} \equiv \frac{v_{i}(0) \, 
\dot{{\rm {se}}}_{\nu}\left(\delta_{0},-\theta\right) \, - \,  
\dot{v}_{i}(0) \,
 {\rm {se}}_{\nu}\left(\delta_{0},-\theta\right)
}{\left[{\rm {ce}}_{\nu}\left(\delta_{0},-\theta\right) \, 
 \dot{{\rm {se}}}_{\nu}\left(\delta_{0},-\theta\right) \, - \,
\dot{{\rm {ce}}}_{\nu}\left(\delta_{0},-\theta\right) \,
 {\rm {se}}_{\nu}\left(\delta_{0},-\theta\right)
\right]}}}, \nonumber \\
{\displaystyle{ 
B_{i} \equiv  -\frac{v_{i}(0) \, 
\dot{{\rm {ce}}}_{\nu}\left(\delta_{0},-\theta\right) \, - \,  
\dot{v}_{i}(0) \,
 {\rm {ce}}_{\nu}\left(\delta_{0},-\theta\right)
}{\left[{\rm {ce}}_{\nu}\left(\delta_{0},-\theta\right) \, 
 \dot{{\rm {se}}}_{\nu}\left(\delta_{0},-\theta\right) \, - \,
\dot{{\rm {ce}}}_{\nu}\left(\delta_{0},-\theta\right) \,
 {\rm {se}}_{\nu}\left(\delta_{0},-\theta\right)
\right]}}},
\end{array}
\label{ecua34}
\end{equation}
\begin{equation}
\begin{array}{l}
{\displaystyle{
W = -\sqrt{l_{s}} \dot{v}_{2}(0)}},    
\end{array}
\label{ecua35}
\end{equation}
and the characteristic curve, Eq.\ (\ref{ecua31}) ,
for our initial parameters can be written  
up to order $\epsilon^{2}$ as  
\begin{equation}
\begin{array}{c}
{\displaystyle{
\delta = \frac{\Omega_{R}}{2 \nu} - 
\frac{q^{2} \nu \cos(2 \delta_{0})}{2 M_{0}^{2} \gamma_{0}^2 \Omega_{R}} 
\epsilon^{2} + O(\epsilon^{4})}}.  
\end{array}
\label{ecua36}
\end{equation}
To obtain a better approximation, we need to take into account 
more  terms of the above series, so in  Fig.\ \ref{f2} we have plotted,  
as solid lines, the characteristic curves
obtained numerically 
with {\em Mathematica 3.0} \cite{Math},  
for $\nu=1/2$ and $\nu=3/2$ 
when $u(0)=0$ and $\delta_{0}=\pi/2$. 
Notice that when $\nu=m+p/s$ is rational, 
with $m$ an integer number and 
$p/s$ a rational fraction ($0 < p/s < 1$),  
 $v_{1}(t)$ or $v_{2}(t)$ are $2 \pi s$ periodic functions, if $p$ is odd,  
and $\pi s$-periodic functions, if $p$ is even; 
whereas for irrational $\nu$ both functions will 
be non-periodic, but bounded solutions \cite{Mathieu}. 

If we know $l(t)$, then from the solution 
$P(t)$ [Eq.\ (\ref{ecua4})], 
we can calculate the velocity  ${\displaystyle{ 
\dot{X}=P(t) l(t)/(M_{0} l_{0})}}$ for the kink center. 
Since the momentum $P(t)$ is a periodic function and the kink
width $l(t)$ at least is a bounded function, 
$\dot{X}$ is bounded as well.
For instance, if we take $\nu = 1/2$ in (\ref{ecua36}), 
the frequency of the external force   
$${\displaystyle{\delta = \Omega_{R} - 
\frac{q^{2} \cos(2 \delta_{0})}{4 M_{0}^{2} 
\gamma_{0}^2 \Omega_{R}} \epsilon^{2} 
+ O(\epsilon^{4})}}, 
$$ 
is very close to $\Omega_{R} \approx \Omega_{i}$ if $\epsilon \ll 1$  
(see Fig.\ \ref{f2}), and $l(t)$ is a function of the square of 
the $4 \pi/\delta$-periodic solutions  
 ${\rm se}_{1/2}$ and ${\rm ce}_{1/2}$, respectively, so the velocity of the 
kink center, ${\dot{X}}(t)$, and the energy $E(t)$ 
will also be bounded functions for $\delta=\delta(\epsilon)$.  

If we try to find, e.g., $2 \pi$-periodic solutions of (\ref{ecua28}) 
we obtain that in this 
case the two independent solutions are related to the integer Mathieu
functions ${\rm se}_{1}$ and ${\rm ce}_{1}$. 
However, these two solutions appear when the characteristic curve 
$a(\theta)$ is 
$$
{\displaystyle{
a = 1 + \theta - \frac{\theta^{2}}{8} - \frac{\theta^{3}}{64} - 
\frac{\theta^{4}}{1536} + ...}},
$$ 
for the even Mathieu function ${\rm ce}_{1}$ and 
$$
{\displaystyle{
a = 1 - \theta - \frac{\theta^{2}}{8} + \frac{\theta^{3}}{64} - 
\frac{\theta^{4}}{1536} + ...}},  
$$
for the odd Mathieu function ${\rm se}_{1}$. Since the values 
of $a$ are different for each of the 
function ${\rm se}_{1}$ and ${\rm ce}_{1}$, 
these functions are not solutions of the same equation except when 
$\theta=0$. Although ${\rm se}_{1}$ and ${\rm ce}_{1}$ are not solutions of 
Eq.\ (\ref{ecua28}) [recall that in order to find $l(t)$ we need to
calculate two independent solutions of Eq.\ (\ref{ecua28})], the characteristic
curves of these solutions separate 
the unstable and stable solutions of the Mathieu equation (\ref{ecua28}), and 
can be rewritten as   
\begin{equation}
\begin{array}{l}
{\displaystyle{
\delta_{-} = \frac{\Omega_{R}}{2} - \left[
\frac{q^2 \cos(2 \delta_{0})}{2 M_{0}^2 \gamma_{0}^2 \Omega_{R}} + 
\frac{q^2}{4 M_{0}^2 \gamma_{0}^2 \Omega_{R}} \right] \epsilon^2}},  
\end{array}
\label{ecua37}
\end{equation}
and 
\begin{equation}
\begin{array}{l}
{\displaystyle{
\delta_{+} = \frac{\Omega_{R}}{2} - \left[ 
\frac{q^2 \cos(2 \delta_{0})}{2 M_{0}^2 \gamma_{0}^2 \Omega_{R}} -  
 \frac{q^2}{4 M_{0}^2 \gamma_{0}^2 \Omega_{R}} \right] \epsilon^2}}, 
\end{array}
\label{ecua38}
\end{equation} 
respectively. 
Interestingly, we note that 
the width of the unbounded region, 
$$\Delta \delta \equiv \delta_{+} - \delta_{-} = 
{\displaystyle{ \frac{q^2 \epsilon^2}{2 M_{0}^2 \gamma_{0}^2 \Omega_{R}} },}$$
decreases when the initial velocity increases, and consequently we will focus
on the case $u(0)=0$ in our numerical simulations below, as the resonance
is then easier to observe.  
The curves $\delta_{+}$ and $\delta_{-}$ [Eqs.\ (\ref{ecua37}) and 
(\ref{ecua38})]  
are plotted in Fig.\ \ref{f2} 
for zero initial velocity and $\delta_{0}=\pi/2$, 
where the shadowed region represents the region where 
$l(t)$ is unbounded. Analogously, we can obtain other
characteristic curves, related 
either with integer Mathieu functions  ${\rm se}_{n}$ and ${\rm ce}_{n}$ 
($n \in {\sf N}$), or with non-integer Mathieu functions 
${\rm se}_{\nu}$ and ${\rm ce}_{\nu}$. 
In view of this, the above analytical results lead us to expect that if  
 $\delta$ is close to $\Omega_{R} \approx \Omega_{i}$;  
$l(t)$, $u(t)$ and the energy should be oscillatory (or, at least,  
bounded) functions, whereas 
if ${\displaystyle{\delta \approx \Omega_{R}/2 \approx 
\Omega_{i}/2}}$ the kink width $l(t)$ increases indefinitely and 
since the energy and the velocity are proportional to $l(t)$ [see Eqs. 
(\ref{ecua4}) and (\ref{ecua22})] we should observe that these 
functions increase with time as well. 

\subsection{ $\lambda \ne 0$ : Resonances at $\delta \approx \Omega_{R}/2$, 
$ \Omega_{R} $ }

When $\lambda \ne 0$, the linear part of Eq.\ (\ref{ecua26}) 
becomes,
in a manner similar to the procedure leading to Eq.\ (\ref{ecua28}),
a more general Mathieu equation \cite{ww}, namely 
\begin{equation}
\begin{array}{l}
{\displaystyle{
g'' + \left[b + 2 \theta_{1} \epsilon^2 \cos(2 \tau) +
 2 \theta_{2} \epsilon^2 \cos(4 \tau) \right] g = 0}},  
\end{array}
\label{ecua39}
\end{equation}
where 
$$
{\displaystyle{
b=\left(\frac{\Omega}{\delta}\right)^2  \left[1 + \frac{q^2 \epsilon^2}
{2 \delta^2 M_{0}^2} + \frac{\lambda^2}{M_{0}^2} \right]}},
$$
$$
{\displaystyle{
\theta_{1} = \left(\frac{\Omega}{\delta}\right)^2 
\frac{\lambda q^{2}}{\delta^{2} M_{0}^2}, \,\ \,
\theta_{2} = \left(\frac{\Omega}{\delta}\right)^2 
\frac{q^{2} }{4 \delta^{2} M_{0}^2}}}, \,\ \,\  
$$
and the prime denotes the derivative 
with respect to $\tau=(\delta t + \delta_{0})/2$. 
According to Floquet 
theory \cite{mook}, Eq.\ (\ref{ecua39}) has normal solutions of the form 
$g=\exp(\sigma \tau) \Phi(\tau)$, 
where $\Phi(\tau)$ is a $\pi$-periodic function 
and $\sigma$ is the characteristic exponent. 
Expanding $\Phi(\tau)$ in a Fourier series, the function $g$ 
can be rewritten as 
$g=\sum_{n=-\infty}^{+\infty} \Phi_{n} \exp[(\sigma + 2 n i) \tau]$, where 
$\Phi_{n}$ are the coefficients of the above series. 
Therefore, the transition curves separating stability from 
instability correspond to $\sigma=0$ ($\pi$-periodic solutions) and 
$\sigma=i$ ($2 \pi$-periodic solutions) \cite{bateman}. 
Now we will apply the method of the strained parameters \cite{mook} to 
determine the values of $b$, $\theta_{1}$ and $\theta_{2}$ corresponding to 
these values of the characteristic exponent. First we assume the solutions 
of (\ref{ecua39}), which have period $\pi$ or $2 \pi$ 
and the transition curves $b=b(\epsilon)$, 
can be written 
in the form of perturbation expansions as
\begin{equation}
\begin{array}{l}
{\displaystyle{  
g(\tau) = g_{0} + \epsilon^2 g_{1} + \epsilon^4 g_{2} + O(\epsilon^6)}}, 
\,\ \nonumber \\  
{\displaystyle{ 
b = b_{0} + \epsilon^2 b_{1} + \epsilon^4 b_{2} + O(\epsilon^6)}}.  
\end{array}
\label{ecua40}
\end{equation}
Second, substituting (\ref{ecua40}) into (\ref{ecua39}) and equating to zero 
the coefficients of the same order of $\epsilon$ we obtain 
\begin{equation}
\begin{array}{l} 
{\displaystyle{  
{\cal L} g_{0} = 0,}}
\end{array}
\label{ecua41}
\end{equation}
\begin{equation}
\begin{array}{l}
{\displaystyle{ 
{\cal L} g_{1} = - b_{1} g_{0} - 2 \theta_{1} \cos(2 \tau) g_{0} - 
2 \theta_{2} \cos(4 \tau) g_{0}}}, 
\end{array}
\label{ecua42}
\end{equation}
\begin{equation}
\begin{array}{l}
{\displaystyle{ 
{\cal L} g_{2}  =  - b_{1} g_{1} - b_{2} g_{0} - 
2 \theta_{1} \cos(2 \tau) g_{1} - 
2 \theta_{2} \cos(4 \tau) g_{1}}},  
\end{array}
\label{ecua43}
\end{equation}
where ${\cal L}$ represent the  
second-order linear differential operator ${\displaystyle{ 
{\cal L} = \frac{d^2}{d \tau^2} + b_{0}}}$. 
The solution of Eq.\ (\ref{ecua41}) is 
\begin{equation}
\begin{array}{l}
{\displaystyle{  
g_{0} = A \cos\left(\sqrt{b_{0}} \tau\right) + 
B \sin\left(\sqrt{b_{0}} \tau\right)}}, 
\end{array}
\label{ecua44}
\end{equation}
where $b_{0}=4 n^2$ for the $\pi$-periodic solutions and  $b_{0}=(2 n -1)^2$ 
for the $2 \pi$-periodic solutions, $n$ is an integer number and $A$ and $B$ 
are constants. 

For the  $\pi$-periodic solutions we will analyze the case $n=1$, 
corresponding to $b_{0}=4$; we note that
the case $n=0$ is not possible due to the definition 
of our parameter b. When $b_{0}=4$, $g_{0}= A \cos(2 \tau) + B \sin(2 \tau)$. 
For $g_{1}$ to be a $\pi$-periodic function is necessary to eliminate 
the secular terms in Eq.\ (\ref{ecua42}), imposing either 
$b_{1} = \theta_{2}$ and $A=0$ or $b_{1} = - \theta_{2}$ and $B=0$. Then, the 
transition curves corresponding to the solutions 
\begin{equation}
\begin{array}{l}
{\displaystyle{  
g_{+} = B \sin(2 \tau) + B \epsilon^2 \left[
\frac{\theta_{1}}{20} \sin(4 \tau) + \frac{\theta_{2}}{32} \sin(6 \tau) 
\right]}}, 
\end{array}
\label{ecua45}
\end{equation}
and 
\begin{equation}
\begin{array}{l}
{\displaystyle{
g_{-} = A \cos(2 \tau) + A \epsilon^2 \left[ \frac{-\theta_{1}}{4} + 
\frac{\theta_{1}}{12} \cos(4 \tau) + \frac{\theta_{2}}{32} \cos(6 \tau) 
\right]}},
\end{array}
\label{ecua46}
\end{equation} 
can be shown to be 
$$
{\displaystyle{
b = 4 \pm \epsilon^2 \theta_{2}}}, 
$$
or, alternatively,
\begin{eqnarray}
\label{ecua47}
\delta_{+} & = & \frac{\Omega_{R}}{2} - 
\frac{q u(0) \cos(\delta_{0})}{M_{0} \gamma_{0}} \epsilon + 
\left[ \frac{3 q^2 }{4 M_{0}^2 \gamma_{0}^2 \Omega_{R}} + 
\frac{q^2 \cos(2 \delta_{0}) }{2 M_{0}^2 \gamma_{0}^2 \Omega_{R}} 
\right] \epsilon^2,  \\ 
\delta_{-} & = & \frac{\Omega_{R}}{2} 
- \frac{q u(0)\cos(\delta_{0}) }{M_{0} \gamma_{0}} \epsilon 
+ \left[ \frac{5 q^2}{4 M_{0}^2 \gamma_{0}^2 \Omega_{R} } + 
\frac{q^2 \cos(2 \delta_{0}) }{2 M_{0}^2 \gamma_{0}^2 \Omega_{R}} 
\right] \epsilon^2. \nonumber 
\end{eqnarray}    
respectively. Since these curves start from $\epsilon=0$ and $b=4$, we are 
analyzing the transition curves when the frequency of 
the ac force is close 
to half of $\Omega_{R}$. Hence, the resonance at 
${\displaystyle{ \delta \approx \Omega_{i}/2 \approx \Omega_{R}/2}}$ 
appears much as in the case $\lambda = 0$ even when $\lambda \ne 0$.
Let us remark that,
for given $\epsilon$, the distance between these two curves 
$\delta = \delta(\epsilon)$ is
$${\displaystyle{ \frac{q^2 \epsilon^2}{2 M_{0}^2 \gamma_{0}^2 \Omega_{R}}},}$$ 
and when the velocity increases, the unstable region is more narrow, 
as we found above in the case when $\lambda = 0$. 

The other transition curves are related to the 
$2 \pi$-periodic solution of Eq.\ (\ref{ecua41}), 
$g_{0} = A \cos(\tau) + B \sin(\tau)$, which appear  
when $b_{0}=1$ ($n=1$). In this case $g_{1}$ will be a
periodic function if $b_{1}=\theta_{1}$ and $A=0$ or 
$b_{1}=-\theta_{1}$ and $B=0$. Hence, the transition curves starting from 
$b=1$ are 
$$
{\displaystyle{
b = 1 \pm \epsilon^2 \theta_{1}}}, 
$$
or, equivalently, 
\begin{equation}
\begin{array}{l}
{\displaystyle{
\delta_{\mp} = \Omega_{R} 
- \frac{q u(0) \cos(\delta_{0}) }{M_{0} \gamma_{0} } \epsilon
+ \left[ \frac{q^2 }{2 M_{0}^2  \gamma_{0}^2 \Omega_{R} } + 
 \frac{q^2 \cos(2 \delta_{0}) }{4 M_{0}^2 \gamma_{0}^2  \Omega_{R}} 
\right] \epsilon^2 \mp 
 \frac{q^2 \lambda} {2 M_{0}^2 \gamma_{0}^2 \Omega_{R} } \epsilon^2}}, 
\end{array}
\label{ecua48}
\end{equation}
and correspond to the solutions 
\begin{equation}
\begin{array}{l}
{\displaystyle{
g_{-} = B \sin(\tau) + B \epsilon^2 
\left[ \frac{\theta_{1}-\theta_{2}}{8} 
\sin(3 \tau) + \frac{\theta_{2}}{24} \sin(5 \tau) 
\right]}},  
\end{array}
\label{ecua49}
\end{equation}
\begin{equation}
\begin{array}{l}
{\displaystyle{
g_{+} = A \cos(\tau) + A \epsilon^2 
\left[ \frac{\theta_{1}+\theta_{2}}{8} 
\cos(3 \tau) + \frac{\theta_{2}}{24} \cos(5 \tau) \right]}}, 
\end{array}
\label{ecua50}
\end{equation}
respectively. These solutions and its corresponding 
characteristic curves are related with the driving frequency 
$\delta$ close to $\Omega_{R}$. Finally, in this case we have
$${\displaystyle{ 
\Delta \delta \equiv \delta_{+} - \delta_{-} = 
\frac{q^2 \lambda \epsilon^2 }{M_{0}^2 \gamma_{0}^2 \Omega_{R}}},}$$ which  
means as above
that for fixed $\lambda$, $\Delta \delta$ decreases when $u(0)$ is 
increased. 

In summary, 
we have found here that when $\lambda \ne 0$ and we drive 
the system with a frequency close to $\Omega_{R}$ 
the solutions will be unstable 
in the region between the curves $\delta_{+}$ and $\delta_{-}$. 
Notice that for $\lambda = 0$ we recover the previous results, i.e., 
there are not any resonances at $\delta \approx \Omega_{R}$. 

\section{Numerical verification}
  
The results we have obtained in the previous section have been derived
within the collective coordinate assumption that we can describe all 
the kink dynamics by the two variables $X(t)$ and $l(t)$, all other
degrees of freedom being negligible. As there is no way to know {\em 
a priori} that this is indeed the case, we have to verify this 
hypothesis by numerical simulations of the full partial differential 
equation. Furthermore, we have not been able to solve the collective 
coordinate equations for the damped ($\beta\neq 0$) case; it is 
reasonable to expect that in this situation phenomena similar to those
found for the undamped case will arise, but in so far this is not 
checked this assertion remains a conjecture. 

In view of the above considerations, we
have computed the numerical solution of the partial differential
equations (\ref{ecua6}) and
(\ref{ecua7}) by using the Strauss-V\'azquez scheme
\cite{stvaz} and choosing a total length for our numerical system of 
$L=400$, with steps
$\Delta t =0.01$, $\Delta x = 0.1$, free boundary conditions, 
and simulating up to a final time equal to $25 \, 000$ with 
a kink at rest as initial condition. 
In addition, we have fixed the amplitude 
$\epsilon = 0.01$ and the phase $\delta_{0} = \pi/2$ of the ac force and 
then we have changed 
the value of the driving frequency $\delta$ in order to see the 
resonances. We note that there are many other parameters we could change,
such as the initial velocity, the driving amplitude,  or the driving phase,
but as our main goal in this section is to assert the validity of the 
general analytical results obtained above we prefer to concentrate on  
a few cases as mentioned. In all our simulations we have chosen the 
initial parameters in such a way that $\lambda = 0$, because when 
$\lambda \ne 0$ the kink moves to the right or to the left and then for 
large times the kink will leave our finite system. 
For the aforesaid values of initial parameters 
the width of the resonance region 
$\Delta \delta = \delta_{+} - \delta_{-}$ 
predicted at $\delta \approx \Omega_{R}/2 \approx \Omega_{i}/2$ in 
the section III (A) [see Eqs.\ (\ref{ecua37}) and (\ref{ecua38})],  
is of the order of $10^{-4}$. For this reason we have explored the 
regions around $\Omega_{i}$ and 
$\Omega_{i}/2$ in an interval of that order. 
Finally, we want to mention that 
with the Strauss-V\'azquez method one can compute very accurately 
the position and the velocity of the kink center using the integrals
of energy and momentum \cite{jim}, so it is a good numerical method
in order to compare to our analytical predictions. 
  
\begin{figure}
\begin{tabular}{c}
\epsfig{file=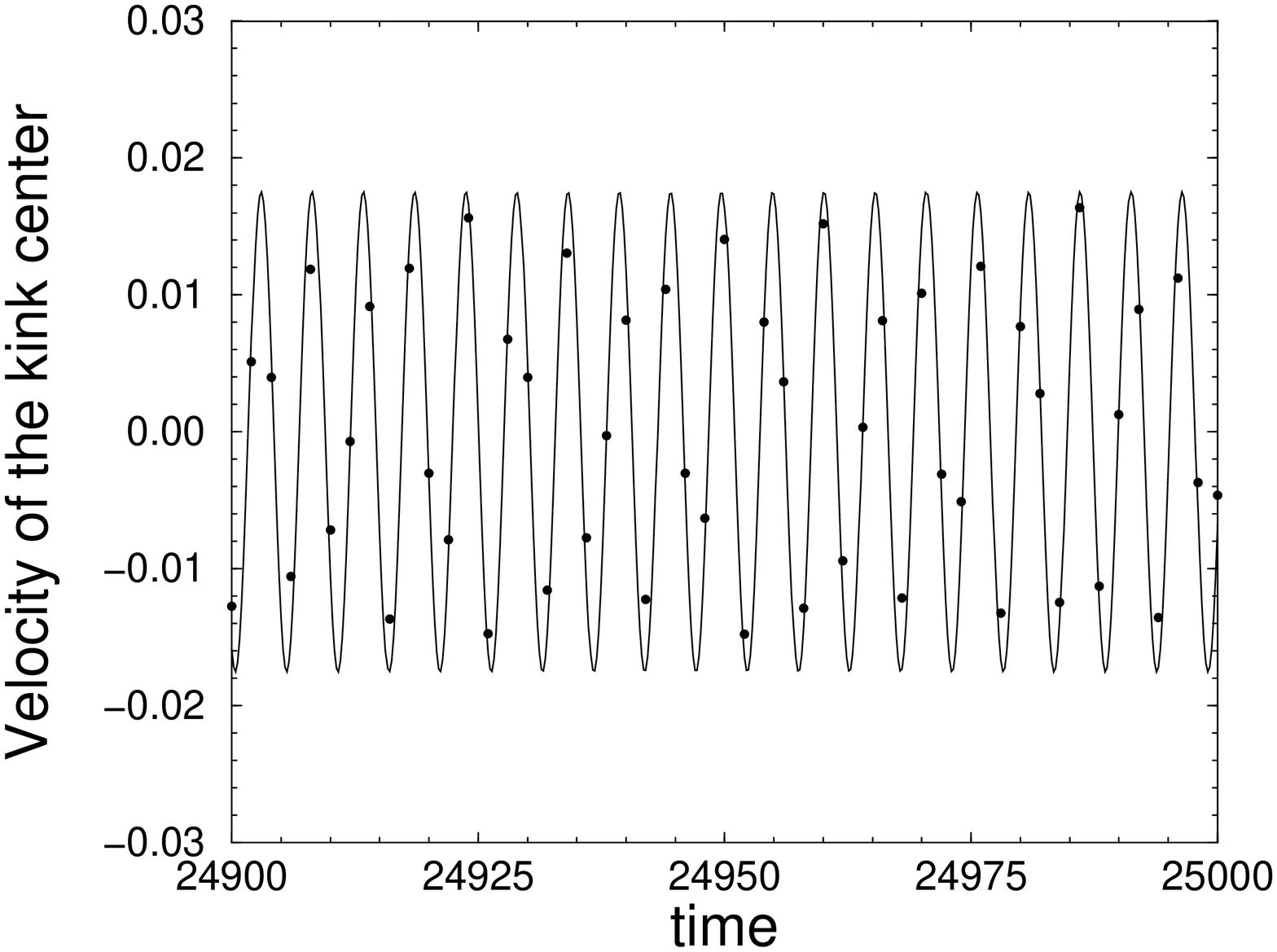, width=3.0in, angle=-90}
\end{tabular}
\caption{Verification of the collective coordinate method in the 
absence of damping. 
The solid line has been obtained from Eqs.\ (\ref{ecua23}) and 
(\ref{ecua24}) and represents the values of the 
velocity of the kink center as given by the collective coordinate 
approach; 
the points are the velocities  
of the center of the kink from the numerical simulations of Eqs.\ (\ref{ecua6})-(\ref{ecua7}) 
by using the  Strauss-V\'azquez scheme, starting from a kink at rest and with 
force $f(t)=0.01 \sin(1.21 t + \pi/2)$. The velocity function is 
represented only from $t=24\,900$ to the final time $t=25\,000$, 
but the same behaviour of this 
function is observed during the whole run.
}
\label{f3}
\end{figure}

First of all, 
in order to verify the results obtained by 
means of the GTWA with two independent collective coordinates, 
the position and the width of the kink,    
we have studied the region around 
$\delta \simeq \Omega_{i} = 1.2247$, i.e.,
the driving frequency 
for which we do not expect any resonances when $\lambda=0$; 
hence, the width and the velocity should be, at least, 
bounded functions.
We have verified that this prediction is in excellent
agreement with the numerical simulations, an example of which is shown in
Fig.\ \ref{f3}. In this plot, 
we can see that the velocity  
is indeed an oscillatory function and even for large times our theoretical 
approach describes correctly the evolution of the velocity of 
the center of mass of the kink. Other values close to $\delta=1.21$
behave very much like the presented example. 
 
\begin{figure}
\begin{tabular}{c}
\epsfig{file=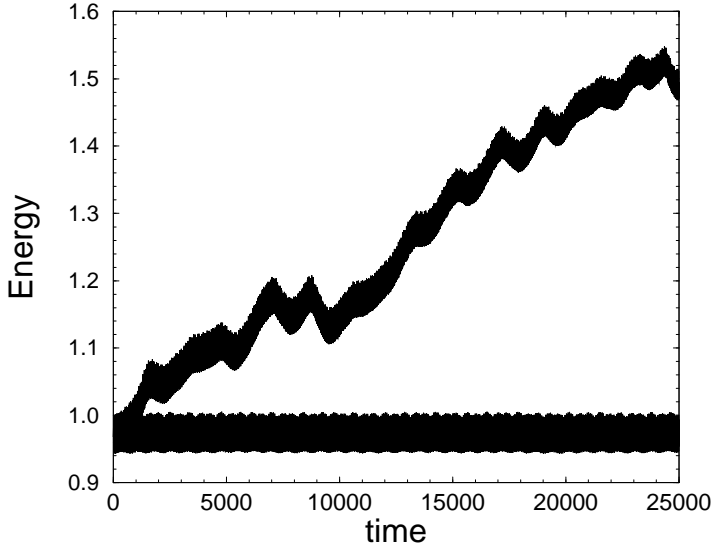, width=3.0in, angle=-90}
\end{tabular}
\caption{Kink energy when  
the frequency of the driving force $\delta$ is close to $\Omega_{i}/2$. 
The 
parameters of the numerical simulations of the full partial differential
equations, 
Eqs.\ (\ref{ecua6})-(\ref{ecua7}), are $\beta=0$, $\epsilon=0.01$, 
$\delta_{0}=\pi/2$, $u(0)=0$, 
$\delta = 0.6104$ (upper curve, resonantly increasing energy)
and $\delta = 0.608$ (lower curve, non-resonant oscillations), 
$L=400$, $\Delta x=0.1$, $\Delta t=0.01$. Notice that half of 
the Rice frequency is $\Omega_{R}/2=0.6226$ and half of the frequency of 
the internal mode is $\Omega_{i}/2=0.6124$.}
\label{f1}
\end{figure}

\begin{figure}
\epsfig{file=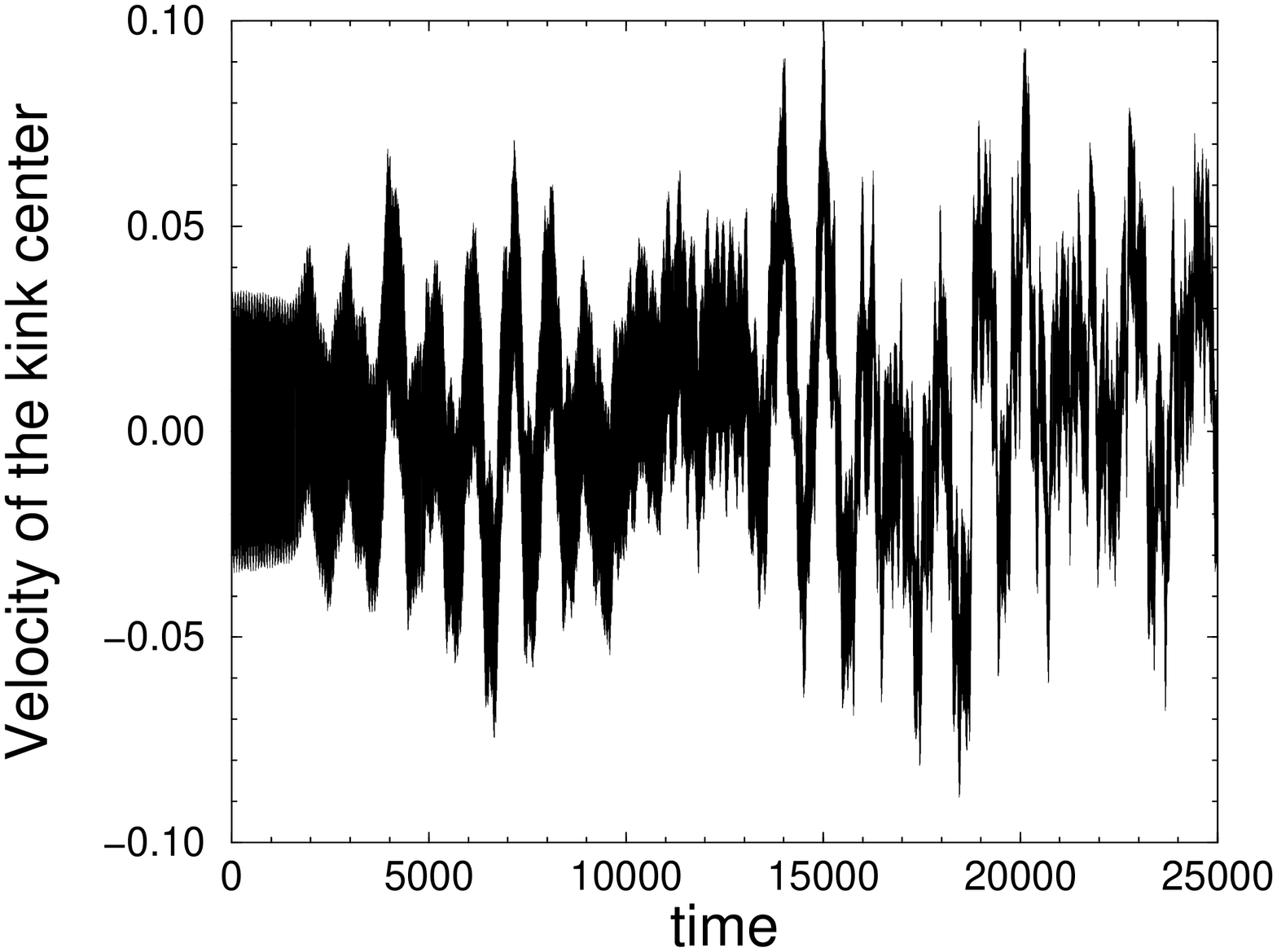, width=2.5in, angle=-90}

\epsfig{file=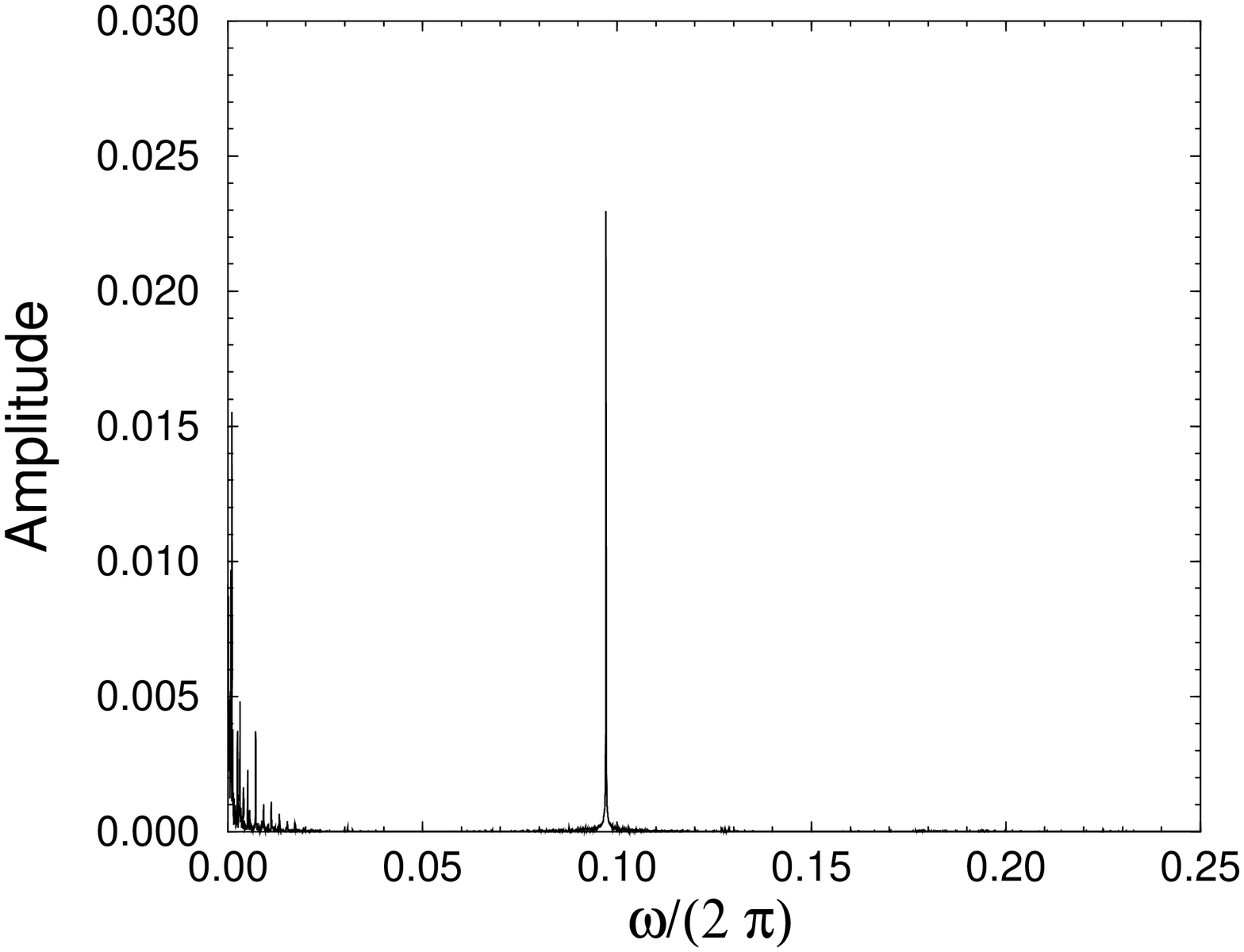, width=2.5in, angle=-90}
\caption{Upper panel: Velocity of the kink center computed 
from the numerical solution of the partial differential 
equation when the frequency of the driving is close to half of 
the internal mode frequency ($\delta=0.6104$), $\beta=0$, 
$\epsilon=0.01$, $\delta_{0}=\pi/2$ and $u(0)=0$. 
After some time the velocity function departs from its oscillatory 
behavior, which was transiently exhibited at the beginning of the run,
and develops chaotic features.
Lower panel: Discrete Fourier transform of the signal in the upper panel.
}
\label{f4}
\end{figure}

The next test of our analytical results has to be, of course, 
the existence of a resonance near $\Omega_i/2$. For frequencies $\delta$
around $\Omega_R/2$, our collective variable approach has predicted that 
the width $l(t)$ increases unboundedly and hence
the velocity and the energy should increase as well. 
Again, the prediction is fulfilled, but for $\Omega_i/2$ 
instead of $\Omega_R/2$: In the numerical simulations
we have observed that in 
this case a resonance takes place   
when $\delta\simeq \Omega_{i}/2 = 0.6124$,   
As a specific example, the results for the kink energy for $\delta=0.6104$ 
and $\delta=0.608$
are plotted in Fig.\ \ref{f1}, clearly showing a resonant increase of the
energy in the former case and oscillatory behavior in the latter. 
Interestingly, at the resonance
the velocity also does not have 
an oscillatory behaviour  
as depicted in Fig.\ \ref{f4}, in contrast to the behavior off-resonance
we found 
in Fig.\ \ref{f3} for $\delta=1.21$. Such a non-periodic evolution of the
kink velocity implies that the kink motion is chaotic as we first found in
\cite{prl} and, eventually, the kink begins emitting radiation. As we 
discussed in \cite{prl}, we believe that this phenomenon has its origin in
the energy transfer from the kink internal mode to the rest of modes in the
system, a mechanism already demonstrated in \cite{campbell,fei}; intuitively,
the reason for that is that the collective coordinate prediction that the
kink width grows without limit can not be physically true due to the own 
nature of the kink, and hence, when the internal mode 
excitation reaches large values the energy ends up being transferred to the
rest of the available modes (translation and radiation). 

\begin{figure}
\begin{tabular}{c}
\epsfig{file=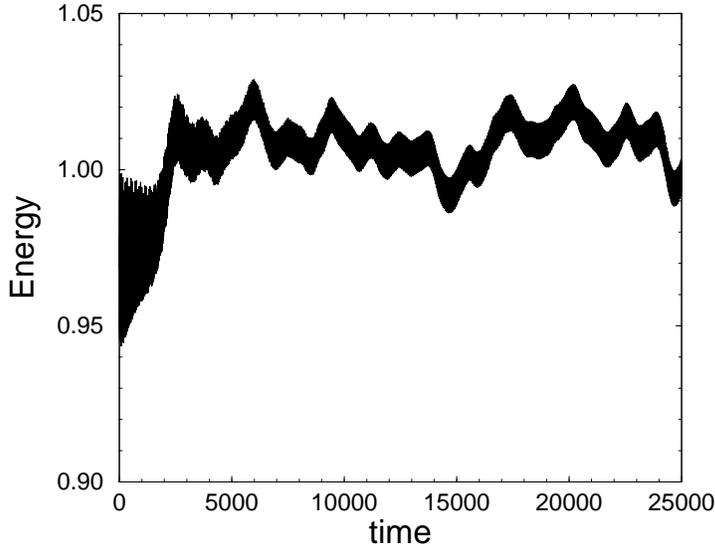, width=3.0in, angle=-90}
\end{tabular}
\caption{Resonance in the presence of damping: Kink energy at
frequency  $\delta=0.6103$ obtained from numerical simulations 
of
Eqs.\ (\ref{ecua6}) and (\ref{ecua7}) with
$\epsilon=0.01$, 
$\beta=0.001$, $u(0)=0$ and $\delta_{0}=\pi/2$. 
}
\label{f5}
\end{figure}

\begin{figure}
\begin{tabular}{c}
\epsfig{file=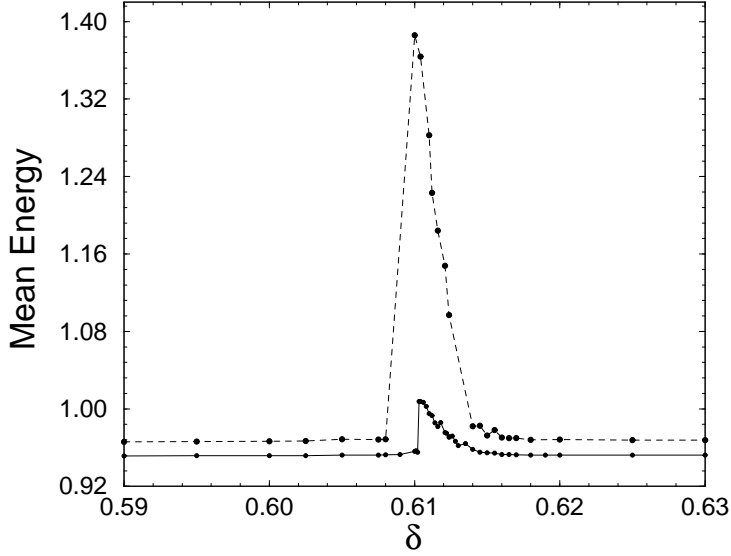, width=3.0in, angle=-90}
\end{tabular}
\caption{Resonance at $\Omega_i/2$. 
Mean energy of the system, computed averaging the energy in
the numerical simulations of the partial differential 
equations for $t>10\,000$ with 
final time equal to $25\,000$; $\epsilon=0.01$, 
$\delta_{0}=\pi/2$, $u(0)=0$. Dashed line: undamped case, $\beta=0$; 
solid line: damped case, $\beta=0.001$.
}
\label{f6}
\end{figure}

\begin{figure}
\begin{tabular}{c}
\epsfig{figure=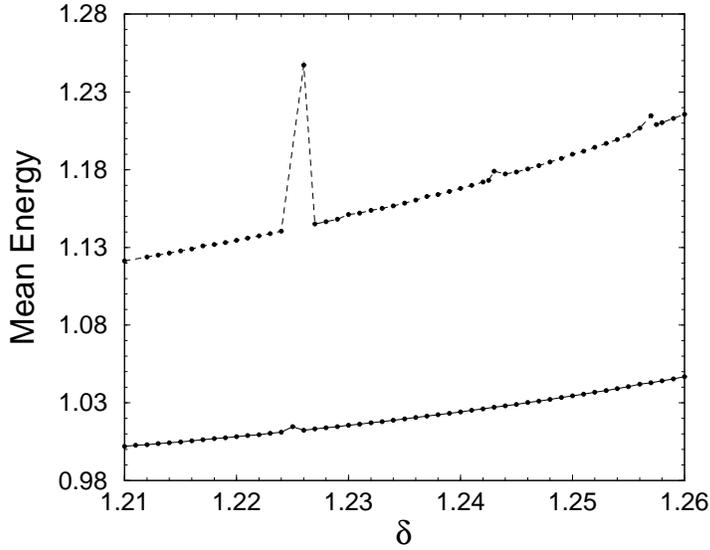,width=3.0in,angle=270}
\end{tabular}
\caption{Resonance at $\Omega_i$. 
Mean energy of the system, computed averaging the energy in
the numerical simulations of the partial differential 
equations for $t>10\,000$ with 
final time equal to $25\,000$; $\epsilon=0.01$,
$\delta_{0}=\pi/2$, $u(0)=0$, $\beta=0$ (dashed line) and
$\beta=0.001$ (solid line).
}
\label{f7}
\end{figure}

After checking our analytical results for the undamped case, we now have
to turn to the damped ($\beta\neq 0$) dynamics, in order to find out 
whether the same phenomena arise there. 
Simulations for the damped case show that the energy 
of the system also increases when the resonance takes place
(see Fig.\ \ref{f5}), but in this case the energy is bounded  
due to the dissipation, reaches an 
asymptotic value, and thus does not 
increase indefinitely as in the absence of 
damping. For 
comparison of both situations 
we have computed the mean energy in the time interval 
$10\,000 < t < 25\,000$ 
with and without damping, and our results are summarized in Figs.\ \ref{f6}
and \ref{f7}. 
For $\beta =0.001$ (see points on the solid line in Fig.\ \ref{f6}) 
we have found that the energy increases at $\delta = 0.6103$, whereas 
for $\beta=0$ (see the points on the dashed line in Fig.\ \ref{f6}), 
the resonance frequency is $\delta \approx 0.6102$. 
   
As mentioned above,
the numerical solutions 
of the full partial differential equation 
show that the 
resonance at $\delta \approx \Omega_{i}/2$ 
also takes place for $\beta \ne 0$, whereas when $\delta \approx \Omega_{i}$ 
(see Fig.\ \ref{f7})
the resonance disappears.
Let us recall that
in all our simulations we have chosen $u(0)=0$ (and hence the 
resonant region should be the widest possible one) and $\delta_{0}=\pi/2$, 
in such a way that $\lambda = 0$. In the case when $\lambda$ vanishes 
we can not expect any resonances 
at $\delta \approx \Omega_{i}$. Nevertheless, in Fig.\ \ref{f7} 
it is clear that 
the energy increases weakly in this region. We believe that 
this maximum 
of the energy is probably due to the fact that the condition for the 
suppression of the resonance
at $\delta \approx \Omega_{i}$ is extremely difficult to 
fulfill numerically. On the other hand, it is also possible that this
condition, which has been 
obtained within the collective coordinate framework,
is only approximately true, and therefore when we choose $\lambda=0$ we 
are beginning with an initial condition close to the one needed for 
complete suppression of the resonance but not exactly there, and hence we
would still see the small bump in Fig.\ \ref{f7}. 
Another possible reason for this peak is that
it is possible that when $\lambda = 0$ at $\delta \approx \Omega_{i} 
\approx \Omega_{R}$ a resonance appears in the next order corrections 
(see the work of Segur \cite{segur}, in which 
the frequency $2 \Omega_{i}$ arises, and could then be parametrically 
excited by $\delta=\Omega_i$). 
Were this the case, this ``weaker 
resonance'', that completely disappears for $\beta \ne 0$ [see the solid line 
in Fig.\ \ref{f7}], is the resonance that we can expect either 
when $\lambda \approx 0$ in the partial differential equation 
or for $\lambda = 0$ and large enough times. 

\section{Conclusions}

We have shown analytically and numerically that 
the internal mode of the $\phi^4$ model can be excited if we drive the system 
with an ac force of a frequency close to $1/2$ of the internal mode 
frequency. This is a very surprising result as the driving we are applying
to the system is not parametric. At resonance,
as a consequence of the increment of the energy 
of the system, the kink initially at rest begins 
to move chaotically \cite{prl} 
and also begins to radiate, i.e., the energy is transferred from the internal 
mode to the translational and radiational ones.
The chaotic motion is confirmed by noting that, when the internal mode 
is excited at  $\delta \approx \Omega_{i}/2$, in the discrete
Fourier transform 
(DFT) of the kink's velocity [see Fig.\ \ref{f4}] we see some 
frequencies in the low frequency part of the 
spectrum aside from the frequency of the ac force,  
$\delta$.
What is more important, 
we have presented a full analysis of the collective coordinate theory
for this problem, whose validity has been undoubtedly confirmed by 
numerical simulations of the full partial differential equation.
In particular, in the absence of damping we have been able to find
analytically all the resonances as well as their dependence on the
kink initial conditions. 
On the other hand, we have also 
shown that the resonance at $\delta \approx \Omega_{i}$ can be 
suppressed if we add a small damping to the system or by an appropriate 
choice of initial parameters of the ac force.
We have verified in our simulations 
that for $\delta \approx \Omega_{i}$ the lower radiational modes  
are excited and for this reason the energy in Fig.\ \ref{f7} increases when 
$\delta$ increases. Of course, this excitation of the lowest phonons 
cannot be explained within the present collective coordinate theory;
a much more involved approach including phonon effects (as in \cite{therm} 
for the sG model) would probably account for that resonance. 

Beyond the application of our results to the $\phi^4$ kink dynamics, 
we believe that the same phenomenology will arise 
in other systems for which internal modes are present as, for example, 
in the double sG equation \cite{kivshar}. 
The analysis we
have presented can be straightforwardly extended to this system, 
and it is quite likely
that similar resonance effects will arise, as well as in other models
with the same feature of internal modes. A related interesting question
concerns the applicability of these results to the sG equation. As we 
mentioned in the introduction, our calculation applies directly to the
sG equation, but in this system the kinks do not posses internal modes. 
Therefore, the identification $\Omega_i\approx\Omega_R$ we have done 
for the $\phi^4$ kink is not available any more, and, what is worse,
the putative $\Omega_R$ lies within the phonon band. However, it has
been reported that close to $\Omega_R$ sG kinks might exhibit a quasi-mode
\cite{willis} with a very long life. If such a mode indeed exists, which 
is still in doubt since no other reports of its existence have been 
published, it could be found by looking at resonances, as we have done
in this paper, at half its frequency, which would be {\em outside}
the phonon band. Work along these lines is in progress \cite{im}.

\section*{Acknowledgments}

We thank Yuri Gaididei, Francisco Dom\'\i nguez-Adame, and 
Jos\'e Cuesta for discussions. 
Work at GISC (Legan\'es) has been supported by 
DGESIC (Spain) grant PB96-0119. Travel between Bayreuth and Madrid has been
supported by ``Acciones Integradas Hispano-Alemanas'', a joint program of
DAAD (Az.\ 314-AI) and DGESIC. 

\appendix

\section{Equivalence of collective coordinate methods}

In this appendix we demostrate that the GTWA, the method we used 
in section II to obtain the collective coordinate equations,
is equivalent to using the variation of the momentum 
and the energy of the system (\ref{ecua1}). First of all, 
from Eqs.\ (\ref{ecua11}) and (\ref{ecua15}) 
we can obtain the variation of the momentum and the energy of the 
perturbed $\phi^{4}$ equation. 
Combining the first three terms of Eq.\ (\ref{ecua11}) and doing some 
straightforward transformations the equation for the momentum can 
be recast in the form 
\begin{equation}
\begin{array}{l}
{\displaystyle{
\frac{dP }{dt} = - \int_{-\infty}^{+\infty} dx 
\frac{\partial \cal H}{\partial X} +  \int_{-\infty}^{+\infty} dx 
F(x,t,\phi,\phi_{t},...) \frac{\partial \phi}{\partial X}}}, \nonumber \\
\nonumber \\ 
{\displaystyle{
P(t) = - \int_{-\infty}^{+\infty} dx \phi_{x} \phi_{t}}}.
\end{array} 
\label{ecua3.1.1}
\end{equation} 
Furthermore, if instead of Eq.\  (\ref{ecua15}) we add 
Eq. (\ref{ecua11}) rewritten as
\begin{eqnarray}
& & \int_{-\infty}^{+\infty} dx \,\ \frac{\partial \phi}{\partial X} 
\frac{\partial \psi}{\partial \dot{X}} \,\ \ddot{X}  +    
\int_{-\infty}^{+\infty} dx \,\  [\phi, \psi] \,\ \dot{l} + 
\int_{-\infty}^{+\infty} dx \,\  \frac{\partial \phi}{\partial X} 
\frac{\partial \psi}{\partial \dot{l}} \,\ \ddot{l} - 
 \nonumber \\
& & - F^{stat} (X)  
 -  \int_{-\infty}^{+\infty} dx F(x,t,\phi,\phi_{t},...) 
\frac{\partial \phi}{\partial X} = 0, 
\nonumber 
\end{eqnarray} 
multiply this equation by $\dot{X}$, and the expression obtained from
(\ref{ecua15}) by $\dot{l}$, we conclude that 
Eq. (\ref{ecua15}) is equivalent to
\begin{eqnarray}
\frac{d H }{dt} & = & \int_{-\infty}^{+\infty} dx 
F(x,t,\phi,\phi_{t},...) \frac{d \phi}{d t}, \nonumber \\ 
 \nonumber \\ 
H(t) &=& \int_{-\infty}^{+\infty} dx \left \{ \frac{\phi_{x}^2}{2} + 
\frac{\phi_{t}^2}{2} + U(\phi) \right \}. 
\label{ecua3.1.2}
\end{eqnarray} 
This implies that when we apply the GTWA in those systems we are, in fact, 
varying the momentum
\begin{eqnarray}
\frac{dP}{dt} & = & - \frac{d}{dt} 
\int_{-\infty}^{+\infty} dx \phi_{x} \phi_{t} = 
- \int_{-\infty}^{+\infty} dx \left[ \phi_{xt} \phi_{t} -  
\phi_{x} \phi_{tt} \right],  
\label{ecua3.1.3}
\end{eqnarray} 
and the energy 
\begin{eqnarray}
\label{ecua3.1.4} 
\frac{ d{\it H}}{dt} & = & \frac{d}{dt} 
\int_{-\infty}^{+\infty} dx \left \{
\frac{1}{2} \phi_{x}^2 + \frac{1}{2} \phi_{t}^2 + U(\phi) \right \} = \\ 
\nonumber  
& = & \int_{-\infty}^{+\infty} dx \left \{
\phi_{x} \phi_{xx} + \phi_{t} \phi_{tt} + U'(\phi) \phi_{t} \right \},   
\end{eqnarray}
where 
\begin{eqnarray}
\phi(x,t)=\phi[x-X(t),l(t)], \,\ \,\ \phi_{t} \equiv \psi(x,t) =  
\psi[x-X(t),l(t),\dot{X}(t),\dot{l}(t)].
\label{ecua3.1.5} 
\end{eqnarray} 

Moreover, we can also start from Eqs.\  (\ref{ecua3.1.3}), (\ref{ecua3.1.4}) 
and (\ref{ecua3.1.5}) and obtain  (\ref{ecua11}) and (\ref{ecua15}). 
Doing that, we only need to take into account that there are 
at least two different ways of transforming the above integrals: 
in one of them one substitutes (\ref{ecua3.1.5}) in 
 (\ref{ecua3.1.3}) and (\ref{ecua3.1.4}), so that  
\begin{eqnarray}
\frac{dP}{dt} & = & 
\int_{-\infty}^{+\infty} dx \,\ \left\{
\frac{\partial \phi}{\partial X} 
\frac{\partial \psi}{\partial \dot{X}} \,\ \ddot{X} +   
 [\phi, \psi] \,\ \dot{l} + 
 \frac{\partial \phi}{\partial X} 
\frac{\partial \psi}{\partial \dot{l}} \,\ \ddot{l}
\right\},  
\label{mom1} \\
\nonumber \\    
\frac{dH}{dt} & = &
\dot{X} \int_{-\infty}^{+\infty} dx \,\ 
\left\{
\frac{\partial \cal{H}}{\partial X} + 
\frac{\partial \phi}{\partial X} 
\frac{\partial \psi}{\partial \dot{X}} \,\ \ddot{X} +   
 [\phi, \psi] \,\ \dot{l} + \frac{\partial \phi}{\partial X} 
\frac{\partial \psi}{\partial \dot{l}} \,\ \ddot{l} \right\} + \nonumber \\
& + & \dot{l} 
\int_{-\infty}^{+\infty} dx \,\ 
\left\{
 [\psi, \phi] \,\ \dot{X}  + 
\frac{\partial \phi}{\partial l} 
\frac{\partial \psi}{\partial \dot{X}} \,\ \ddot{X} + 
\frac{\partial \phi}{\partial l} 
\frac{\partial \psi}{\partial \dot{l}} \,\ \ddot{l} 
\right\},
\label{ener1}  
\end{eqnarray} 
whereas, in the other one, we first substitute the systems of Eqs.\  
(\ref{ecua6})-(\ref{ecua7}) in (\ref{ecua3.1.3}) and 
(\ref{ecua3.1.4}) and then assume that $\phi(x,t)$ and
 $\psi(x,t)$ satisfy (\ref{ecua3.1.5}). Thus,
\begin{eqnarray}
\frac{dP}{dt} = 
-\int_{-\infty}^{+\infty} dx \,\ \frac{\partial {\cal{H}}}{\partial X} + 
\int_{-\infty}^{+\infty} dx F(x,t,\phi,\phi_{t},...) 
\frac{\partial \phi}{\partial X}, 
\label{mom2} \\
\nonumber \\    
\frac{dH}{dt} = \dot{X} \int_{-\infty}^{+\infty} dx \,\ 
F(x,t,\phi,\phi_{t},...) 
\frac{\partial \phi}{\partial X} + \dot{l}    
\int_{-\infty}^{+\infty} dx \,\ F(x,t,\phi,\phi_{t},...) 
\frac{\partial \phi}{\partial l}. 
\label{ener2}  
\end{eqnarray} 
Equating  (\ref{mom1}) and (\ref{mom2}), and 
 (\ref{ener1}) and (\ref{ener2}), respectively, we 
find that (\ref{ecua3.1.3}) and (\ref{ecua3.1.4}) become (\ref{ecua11}) and
(\ref{ecua15}), respectively.

Let us note 
that the connection between the GTWA 
and the the variation of the 
momentum and energy gives a physical interpretation of such a technique, 
and furthermore, this second method leads more 
directly to formulas for $X(t)$ and $l(t)$. 
Furthermore, if the solution of Eqs. (\ref{ecua6})-(\ref{ecua7}) 
for $F(x,t,\phi,\phi_{t},...)= -\beta \phi_{t} + f(x,t)$ is a kink 
centered at  $X(t)$ whose  width, $l(t)$, depends on time, 
(\ref{ecua9})-(\ref{ecua9a}), then Eqs.\ 
(\ref{ecua3.1.3}) and (\ref{ecua3.1.4}) become
\begin{eqnarray}
\frac{dP }{dt} & = & -  \beta P - \int_{-\infty}^{+\infty} dx 
f(x,t) \phi_{x}, \,\ \,\ P(t) = \frac{M_{0} l_{0} \dot{X}(t)}{l(t)};  
\label{ecua3.1.6}
\end{eqnarray} 
and  
\begin{eqnarray}
\label{ecua3.1.7}
& & \frac{P(t) l(t)}{M_{0} l_{0}} \left [ 
\frac{dP}{dt}  + \beta P(t) + \int_{-\infty}^{+\infty} dx 
f(x,t) \phi_{x} \right ] + \\ \nonumber 
& & +
 \dot{l} \left [ 
\frac{P(t)^2 }{2 M_{0} l_{0}} + 
\frac{1}{2} \alpha M_{0} l_{0} \left (\frac{2 \ddot{l}}{l}-
\frac{\dot{l}^2 }{l^2} \right) + \frac{1}{2} M_{0} \left(
\frac{1}{l_{0}}- \frac{l_{0}}{l^2} \right) + 
\right. 
\\ \nonumber 
& & \left.  +
\beta \alpha \frac{l_{0}}{l} \dot{l} +   
\int_{-\infty}^{+\infty} d\theta f(l \theta+X(t),t) \, \theta \phi_{\theta} 
\right] = 0,   
\end{eqnarray}
respectively. 
The first bracket on the right hand of (\ref{ecua3.1.7})  vanishes 
because of (\ref{ecua3.1.6}), so, the solution of (\ref{ecua3.1.7}) is 
$\dot{l}=0$ or   
\begin{eqnarray} 
\label{ecua3.1.8}
\alpha [\dot{l}^{2} - 2 l \ddot{l} - 2 \beta l \dot{l}] & = & 
\frac{l^{2}}{l_{0}^{2}} 
\left [1 +  \frac{P^{2}}{M_{0}^{2}} \right ] - 1 + 
\\ \nonumber 
&+&\frac{2 l(t)^2}{M_{0} l_{0}} \int_{-\infty}^{+\infty} 
d \theta f[\theta l + X(t),t] \, \theta \phi_{\theta}. 
\end{eqnarray} 
In such a way, (\ref{ecua3.1.6}) and (\ref{ecua3.1.8}) can be interpreted 
as the equations which the collective coordinates, $X(t)$ [$P(t)$] 
and  $l(t)$, satisfy, and they are obtained from 
$dP/dt$ and $dH/dt$, respectively. 
For example, if 
$f(x,t)=\epsilon \sin(\delta t +\delta_{0})$ in 
(\ref{ecua3.1.6}) and (\ref{ecua3.1.8}), 
we recover the Eqs.\ (\ref{ecua23}) and (\ref{ecua24}), 
obtained by applying the GTWA in section II. 

As a final remark, we point out that,
since we have not used in any moment the 
explicit form for the potential function $U(\phi)$,  
the equivalence between these two aforesaid methods remains 
true for any other perturbed nonlinear Klein-Gordon equation 
of the form (\ref{ecua6})-(\ref{ecua7}), when the Hamiltonian of the system 
is of the form (\ref{ecua8}).

\end{document}